# Correlation-driven organic 3D topological insulator with relativistic fermions

Tetsuya Nomoto[1,2,*], Shusaku Imajo[1], Hiroki Akutsu[2], Yasuhiro Nakazawa[2], Yoshimitsu Kohama[1]

[1]*The Institute for Solid State Physics, the University of Tokyo, Kashiwa, Chiba 277-8581, Japan*

[2]*Graduate School of Science, Osaka University, Toyonaka, Osaka 560-0043, Japan*

[*]Corresponding author. Email: nomotot21@issp.u-tokyo.ac.jp

**Abstract**

**Exploring new topological phenomena and functionalities induced by strong electron correlation has been a central issue in modern condensed-matter physics. One example is a topological insulator (TI) state and its functionality driven by the Coulomb repulsion rather than a spin-orbit coupling. Here, we report a 'correlation-driven' TI state realized in an organic zero-gap system $\alpha\text{-(BETS)}_2\text{I}_3$. The surface metallic state that emerges at low temperatures exhibits characteristic transport properties of a gapless Dirac semimetal, evidencing the presence of a topological surface state in this compound. Moreover, we observe a topological phase switching between the TI state and non-equilibrium Dirac semimetal state by a dc current, which is a unique functionality of a correlation-driven TI state. Our findings demonstrate that correlation-driven TIs are promising candidates not only for practical electronic devices but also as a field for discovering new topological phenomena and phases.**

**Main**

Topological insulator (TI) is a new class of materials that possess both bulk insulating and exotic surface/edge states [1]. Realizing a TI state usually requires a strong spin-orbit coupling (SOC) of heavy elements, which induces a band inversion and an insulating bulk gap [2]. Thus, it is generally challenging to observe a TI state and its unique physical features in materials composed of light atoms, such as carbon-based systems, because of their weak SOC. However, recent theoretical investigations propose another approach to realize a TI state using a strong electron correlation that the effect of the Coulomb interaction opens the bulk gap [3]. Such a correlation-driven TI is scientifically significant because it broadens the range of materials that can form TI states and has the potential to add new functions unique to correlated systems. Strongly correlated organic conductors with topological band structure are promising candidates for this type of TIs. Here, we report a three-dimensional (3D) TI state realized in a quasi-two-dimensional (quasi-2D) organic conductor α-(BETS)$_2$I$_3$ (Fig.1a) and its anomalous transport properties, where BETS denotes bis(ethylenedithio)tetraselenafulvalene [4].

α-(BETS)$_2$I$_3$ consists of the conducting BETS layers and the non-magnetic insulating I$_3$ layers. The BETS layers accommodate one hole for every two BETS molecules, resulting in a 1/4-filled hole band system. First-principles calculations suggest that the band structure of this compound has a Dirac-cone type linear dispersion with a narrow bandgap of $\Delta$ ~2 meV owing to SOC of Se atoms, and the Fermi energy crosses near the Dirac point (Fig. 1b) [5,6]. α-(BETS)$_2$I$_3$ exhibits a semimetallic behaviour from room temperature and undergoes a metal-insulator (MI) transition at 50 K [7]. A recent theory suggests that the Coulomb interaction enhances the bandgap below 50 K, leading to a robust TI state [8].

Considering the dimensionality of this compound, α-(BETS)$_2$I$_3$ is expected to behave as a stacked 2D-TI (weak 3D-TI) with a gapless state on the edge of the conducting plane (*ab*-plane) (Fig. 1c). However, experimental evidence of the 2D-TI state, such as a quantum spin Hall effect, has not been observed.

We carefully performed electrical conductivity measurements to verify the accurate low-temperature electronic state of α-(BETS)$_2$I$_3$. Fig. 1d shows the temperature dependence of the in-plane (*I* // *ab*-plane) and out-of-plane (*I* // *c*-axis) resistances of α-(BETS)$_2$I$_3$. The in-plane resistance (Sample #1) slightly decreases with decreasing temperature and an exponential increase is observed at the MI transition temperature of 50 K. Furthermore, from 50 to 35 K, $R(T)$ can be fitted well using the Arrhenius equation $R(T) = R_0 \times \exp(\Delta/k_B T)$, with an estimated gap size of $\Delta \sim 30$ meV, which agrees with previous reports [7]. These behaviours are consistent with the 2D-TI picture. However, the slope of the in-plane resistance becomes smaller and saturates at lower temperatures, with a step-like anomaly observed between 35 and 10 K. The saturation of resistance is often observed in a 3D-TI with a surface conduction [9,10], which is a different character from that expected for a 2D-TI. The out-of-plane resistance (Sample #2) is even more anomalous; an abrupt drop at 35 K and a temperature-independent behaviour below 10 K are observed, while a sharp increase is observed at 50 K. Previous bulk measurements, such as the lattice constant measurement, magnetic susceptibility, and nuclear-magnetic-resonance measurements, have not detected any anomalies near 35 K [6,11]. Therefore, the anomaly in the in- and out-of-plane resistances suggest the emergence of surface conduction at 35 K. This indicates that a dimensional crossover from 2D to 3D occurs and the actual low-temperature electronic state of α-(BETS)$_2$I$_3$ is a 3D-TI state (Fig. 1c) despite theoretical predictions [6,8]. The dimensional crossover in TIs has been theoretically predicted, for example, in cold-atomic gases in optical lattices [12], but this is the first experimental observation in solids. The origin of the difference in temperature dependence between the in- and out-of-plane resistances is considered to be the anisotropy of the bulk resistivity of this compound. A detailed discussion on the anisotropy of resistivity, sample dependence, and reproductivity is provided in Supplemental Materials.

The observation of a surface metallic conduction does not necessarily guarantee that this compound has a topological surface state. Thus, we performed magnetoresistance (MR) measurements up to high magnetic fields to reveal the topological character of surface metallic states in α-(BETS)$_2$I$_3$. Because α-(BETS)$_2$I$_3$ has a Dirac-cone type band structure, the carriers on the gapless metallic surface are expected to behave as relativistic Weyl fermions [13]. One of the characteristics of such relativistic fermions is the quantum transport phenomenon caused by the degree of freedom of chirality. In the quantum field theory, massless fermions are divided into left- (clockwise) or right-handed (anti-clockwise) particles. In the absence of any external gauge field, the massless fermions with opposite chirality do not mix with each other, and the balance of chirality is conserved. However, on the application of an electric current (*I*) and magnetic field (*B*) in parallel (*I // B*), the symmetry of the chiral fermions is broken. In general, this results in the Adler-Bell-Jackiw chiral anomaly or chiral magnetic effect (CME) [13-15], where a large negative MR is observed. This CME-induced negative MR is known as critical transport evidence for the presence of relativistic fermions and is observed in inorganic topological semimetals such as Na$_3$Bi [16], Cd$_3$As$_2$ [17], and ZrTe$_5$ [18].

Figure 2a shows the normalised in-plane MR of α-(BETS)$_2$I$_3$ (Sample #1) in the configuration of $I // B$, which results in a negligible effect of the Lorentz force on MR. In the high-temperature region ($T > 10$ K), a positive MR monotonically increases with $B$. However, a dip-like anomaly is observed below 25 T with a further decrease in the temperature ($T < 10$ K). Finally, the dip structure converts into a negative MR below 4 K. At the lowest temperature of the measurement ($T = 1.6$ K), the negative MR reaches -14% at $B = 25$ T. Although a large negative MR is often observed in magnetic molecular compounds owing to electron-spin scattering [19], α-(BETS)$_2$I$_3$ is non-magnetic. Thus, the negative MR in α-(BETS)$_2$I$_3$ can be attributed to CME. In the case of Dirac band systems, CME induces a charge flow in the direction of $I$ (and $B$) to compensate for the chirality imbalance between the two Weyl nodes with distinct chirality $x = +1$ and $x = -1$ (Fig. 2b). In theories of CME [20,21], the positive magnetoconductance (MG) in weak magnetic fields is expressed as $MG \sim MR^{-1} \sim C_a B^2$, where $C_a$ is a fitting parameter. Fig. 2c shows the magnetic-field dependence of MG and results of the quadratic fitting. The consistency of the fitting in a wide magnetic field range ($B < 10$ T) suggests that the positive MG (negative MR) in α-(BETS)$_2$I$_3$ is induced by CME. Above ~25 T, MG exhibits a downturn behaviour. Because the magnetic field was applied parallel to the 2D conduction layers, the effects of the Landau level splitting and quantum limit are considered to be small [20-22]. Thus, the downturn observed in the high-field region is probably caused by the transverse MR or Hall component owing to the slight misalignment of B. Moreover, an additional positive MG of less than 1% in weak magnetic fields ($B < 5$ T) was observed, as shown in the inset of Fig. 2c. Owing to its extremely small value and tendency to appear only at low temperatures and weak magnetic fields, it is considered to be a contribution of weak localization in the bulk insulator region.

Another characteristic magnetotransport property of Dirac systems is the large positive MR effect in the $I \perp B$ configuration [23-25]. Figure 3a shows the magnetic field dependence of MR of Sample #1 when $B$ is perpendicular to the $ab$-plane. In contrast to the $I // B$ configuration, a large and non-saturating positive MR is observed. The positive MR increases with decreasing temperature and exceeds 400% at 2.3 K, 50 T. In addition, it shows $B^2$-dependence at lower magnetic fields ($B < 20$ T) and a sublinear dependence at higher magnetic fields ($B > 20$ T). Such a non-saturating MR is characteristic of compensated metals, such as Dirac semimetals, wherein holes and electrons compensate each other, thereby resulting in a positive MR effect by the Lorentz force that continues to appear in high magnetic fields. Indeed, the positive MR can be qualitatively reproduced using the two-carrier model as well as other topological semimetals[26,27] (see Supplemental Materials). Figure 3b shows the MR of Sample #3 as a function of the elevation angle $\theta$ between $B$ and $E$, measured at 4.3 K. Herein, the negative longitudinal MR by CME is negligibly small compared to the positive transverse MR at this temperature. When $\theta = 0°$, MR exhibits the smallest value owing to the Lorentz force-free configuration. Thereafter, with an increase in $\theta$, MR increases monotonically, reaching the maximum value at $\theta = 90°$. Such a field-direction-sensitive MR is a specific feature universally observed in Dirac/Weyl fermion systems [17,18,23-25]. The relationship between the magnitude and angle of MR (Fig. 3c) can be fitted well with a curve proportional to sin2$\theta$ (the orange dashed line); thus, the origin of the transverse MR in $\alpha$-(BETS)$_2$I$_3$ is the classical Lorentz force. Comparing the MR magnitudes at 2.3 and 100

K for the $I // B$ and $I \perp B$ configurations (Fig. 3d) reveals that no anisotropy of MR can be observed at 100 K (the metallic state). Therefore, the anisotropy of MR originates from the low-temperature surface state rather than the anisotropy of the 2D electronic bulk band structure. Because non-relativistic electrons are produced by a thermal disturbance in the high-temperature metallic state, it is natural that a large positive or negative MR cannot be observed at high temperatures. This is in contrast to that of inorganic topological semimetals, where the CME-induced negative MR can be observed at high temperatures. The temperature dependences of the magnitude of the transverse MR of α-$(BETS)_2I_3$ and the out-of-plane resistance are plotted in Fig. 3e. The temperatures at which the surface conduction emerges and the transverse MR begins to increase agree, suggesting the appearance of an exotic surface state with Dirac-type band dispersion near 35 K. In addition, the CME-induced negative MR is observed below 10 K (Fig. 2a), where the out-of-plane resistance exhibits a temperature-independent behaviour. This indicates that the middle-temperature region between 35 and 10 K is a crossover regime from 2D- to 3D-TI, where massless Weyl fermions are absent. Moreover, no Shubnikov-de Haas oscillations are detected during measurements. It is considered that the absence of quantum oscillations is owing to the coincidence of the Fermi energy with a Dirac point. Indeed, quantum oscillations have been observed on carrier-doped samples in relatively weak magnetic fields below 8 T [28].

The anomalous MR properties reveal that α-$(BETS)_2I_3$ is a correlation-driven 3D-TI. A significant difference from conventional TIs is in that the bulk insulating state is sustained by the electron correlation. It is known that the conventional insulating states derived from the electron correlation, such as Mott insulators and charge-ordered (CO) insulators, can be electrically controlled using a dc current [29,30]. Similarly, a correlation-driven TI state is expected to be controlled by a dc current and be changed into another topological state. Such electrical controllability of topological matters would further broaden practical applications for electronic devices. Here, we challenge to control the correlation-driven TI state of α-(BETS)$_2$I$_3$ by an external dc current.

Figures 4a and 4b show the current-voltage (*I–V*) characteristic and current dependence of the resistance of Sample #4 under constant current conditions, respectively. These measurements were performed under the exchange gas condition using a pulsed current for less than 10 ms to avoid the influence of Joule heating. We observe a giant nonlinear conduction effect at temperatures below the MI transition (Fig. 4a). The sample resistance (*V/I*) decreases with increasing current, and a drastic reduction of approximately four orders of magnitude in resistance is observed at 10 K (Fig. 4b). Such an anomalous *I–V* characteristic is reminiscent of those of a tunnel diode or correlation-driven insulators [29]. The *I–V* curves show a local peak at some voltages ($V_{peak}$); for instance, $V_{peak}$ is 0.072 V (= 7.2 V/cm) at 40 K (Figure S5a). Note that the value of $V_{peak}$ almost corresponds to the threshold voltage $E_{th}$ of nonlinear conduction [30,31]. Based on the Zener breakdown model[32], which is a classical mechanism for nonlinear conduction, $E_{th}$ is estimated as $E_{th} \sim \Delta/ea$, where $\Delta$ is the gap size, $e$ is the elementary charge, and $a$ is a lattice parameter. Using $a \sim 1$ nm from crystallographic data [7] and $\Delta = 30$ meV, $E_{th}$ is estimated to be ~0.3 MV/cm, which is clearly larger than the order of $V_{peak}$. The value of $V_{peak}$ exponentially increases with decreasing temperature, which is observed in the charge-density-wave (CDW) sliding and Mott gap suppression by a current; a thorough discussion is available in Supplemental Materials. The $E_{th}$ value for the depinning of CDW is usually $10^{-1}$–$10^{-2}$ V/cm, which is clearly smaller than the estimated $E_{th}$ value for this compound. Therefore, the origin of the nonlinear conduction of α-(BETS)$_2$I$_3$ is supposed to be the bandgap suppression in the bulk region similar to the nonlinear conduction in Mott insulators. Here, the application of a dc current may reduce the gap size of the correlation-driven TI state, leading to a different non-equilibrium electronic state.

When the bandgap on the Dirac-cone is closed by applying a large current, the bulk Dirac semimetal state likely emerges, which exhibits the anomalous magneto-transport properties caused by relativistic chiral fermions. We investigated the current dependence of MR to clarify the nature of carriers in the possible current-induced Dirac semimetal state in α-(BETS)$_2$I$_3$. Fig. 4c displays the magnetic-field dependence of the longitudinal MR ($I // B$) measured at 10K using the four-terminal ac method with a dc offset current $I_{off}$. When no $I_{off}$ is applied, a positive MR is observed at 10 K, which is consistent with Fig. 2a. As the current is increased, the positive MR is gradually suppressed and turns to negative MR above $I_{off} > 1$ mA (= 27 A/cm$^2$). The positive/negative switching in MR is a unique feature of this material that has not been observed in conventional nonlinear conduction materials[33]. With a large dc current ($I_{off} = 15$ mA), MR exhibits an upturn above approximately 30 T, which is comparable to the longitudinal MR with no offset current at lower temperatures (Fig. 2a). Such current dependence of MR cannot be explained by the Joule heating effect. Furthermore, the CME-induced MR from the surface is negligible because this result was obtained at 10 K. Therefore, the negative MR can be attributed to CME by the current-induced carriers in the bulk. This supports the scenario that the bandgap on the Dirac-cone is forced to be closed by applying a dc current, realizing the non-equilibrium Dirac semimetal state with chiral fermions in α-(BETS)$_2$I$_3$. This suggests that a correlation-driven TI possesses the electrical controllability into a different topological system. This unique feature has not been observed in conventional TIs driven by SOC. Topological materials, whose physical properties can be electrically controlled, are desirable not only as a practical electronic device but also as a system for seeking a novel topological phase and phenomenon. Consequently, our findings demonstrate that correlation-driven TIs hold the potential to be a new playground for the science of topology.

In summary, we revealed that α-(BETS)$_2$I$_3$ is a correlation-driven 3D-TI with relativistic fermions at low temperatures. In addition, we demonstrated the topological phase switching phenomena between the Dirac semimetal and TI by an external current. These findings provide a new direction for the development of topological materials such as TIs and Dirac/Weyl semimetals and for the exploration of their functionality concerning device applications. The origin of the 2D-3D crossover near 35 K and the detailed mechanism of bulk gap closure by applying a dc current are still open questions. The theoretical investigation of the topological characters of α-(BETS)$_2$I$_3$ will be presented in the future study.

**Author contributions**

T. N., S. I., and Y. K. performed the magnetotransport measurements, data analysis, and preparation of the manuscript. T. N., H. A., and Y. N. synthesised the single crystals of α-(BETS)$_2$I$_3$. All authors contributed to the writing of the manuscript.

**Acknowledgement**

This work was supported by a Grant-in-Aid for Scientific Research (Grant No.

**Competing financial interests**

The authors declare no competing financial interests.

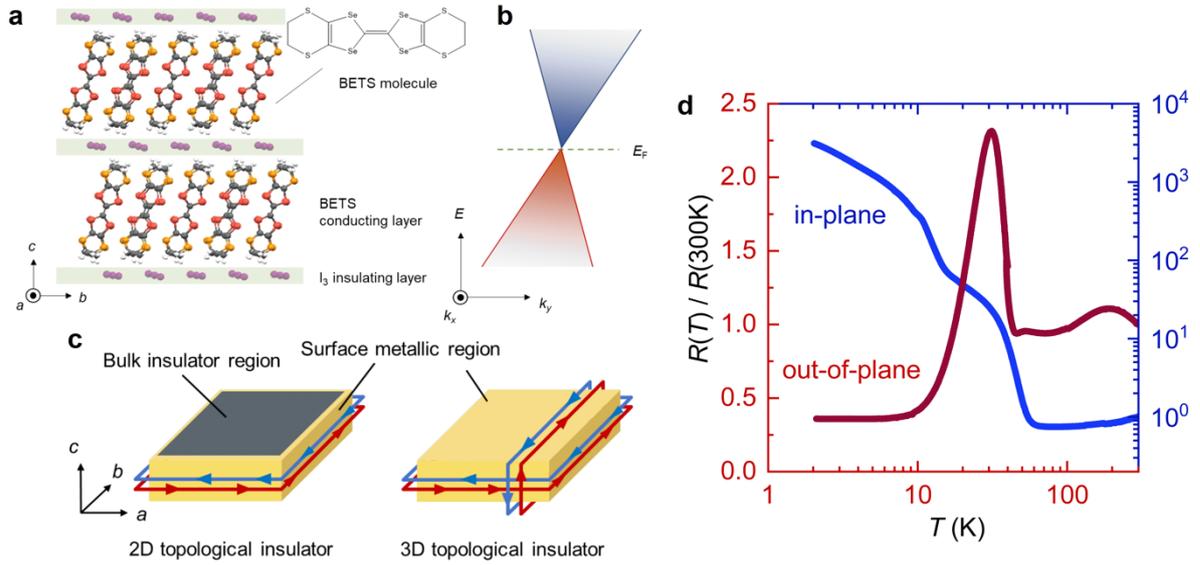

**Figure 1 | Crystal structure and electric properties of α-(BETS)$_2$I$_3$. a**, Side view of the crystal structure of α-(BETS)$_2$I$_3$. BETS molecules forming a 2D-conducting layers (*ab*-plane) are separated by non-magnetic insulating layers of I$_3$. **b**, Schematic of the band structure of α-(BETS)$_2$I$_3$. The valence and conduction bands are touching at the Dirac point and Fermi energy $E_F$ is at the Dirac point. The Dirac-cone is slightly tilted because of the low symmetry of the crystal[9]. **c**, Schematic of 2D- and 3D-TI states in a-(BETS)$_2$I$_3$. In the former, the metallic surface appears on the edge of the insulating *ab*-plane. In the latter, the metallic surface appears on all the crystal faces. **d**, The temperature dependence of the in-plane and out-of-plane resistance. The blue and red curves correspond to the temperature dependence of the in-plane ($I$ // *ab*-plane) resistance (Sample #1) and the out-of-plane ($I$ // *c*-axis) resistance (Sample #2), respectively. Both data are normalized by the value at 300 K. The out-of-plane resistance exhibits a drop at approximately 35 K, indicating the appearance of the metallic surface conduction.

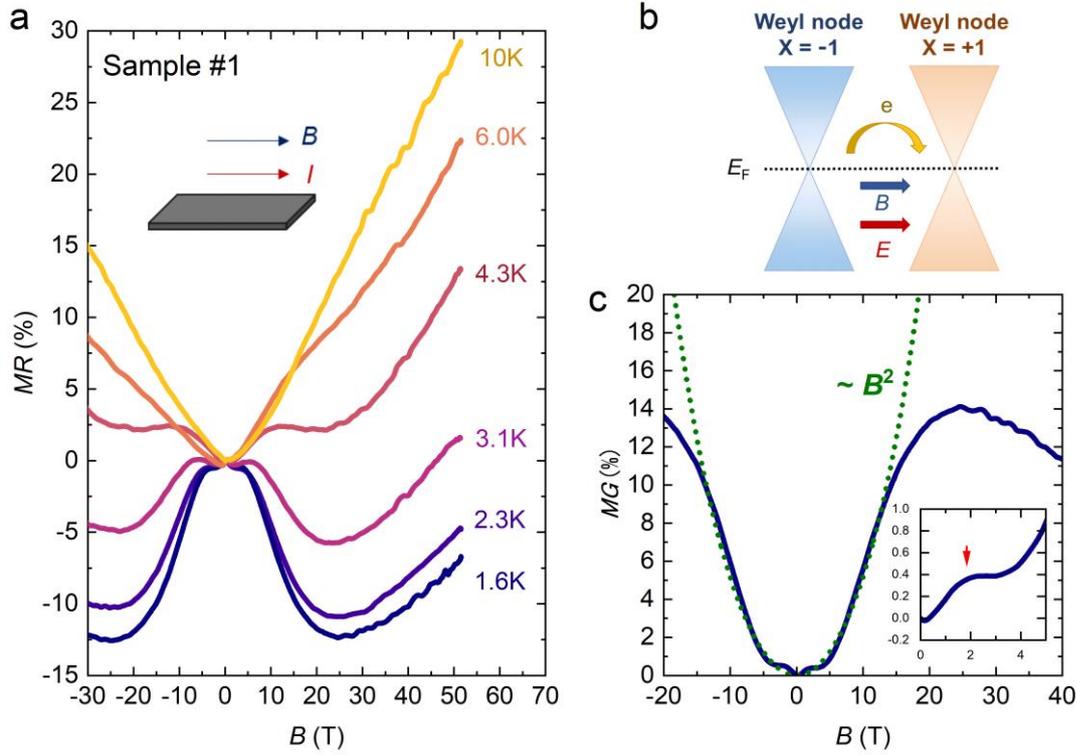

**Figure 2 | Negative magnetoresistance (MR) in α-(BETS)$_2$I$_3$. a,** Longitudinal MR of α-(BETS)$_2$I$_3$ ($I \parallel B$). The magnetic-field dependence of MR measured at several temperatures. The quadratic negative magnetoresistance can be observed below 4 K. **b**, Schematic of the charge flow induced by CME. In a magnetic field, the positive magnetoconductance (MG) due to the chiral anomaly flows between two Weyl nodes with different chirality $x = \pm 1$. **c**, MG at 1.6 K. The green dashed line denotes the fitting curve $MG = C_a B^2$ with $C_a = 4.92 \times 10^{-4}$. Below 15 T, MG follows $B^2$ dependence, indicating the CME-induced MR. Inset: Expanded figure of MG at 1.6 K. The hump-like positive MG can be observed below 5 T, implying the weak localization.

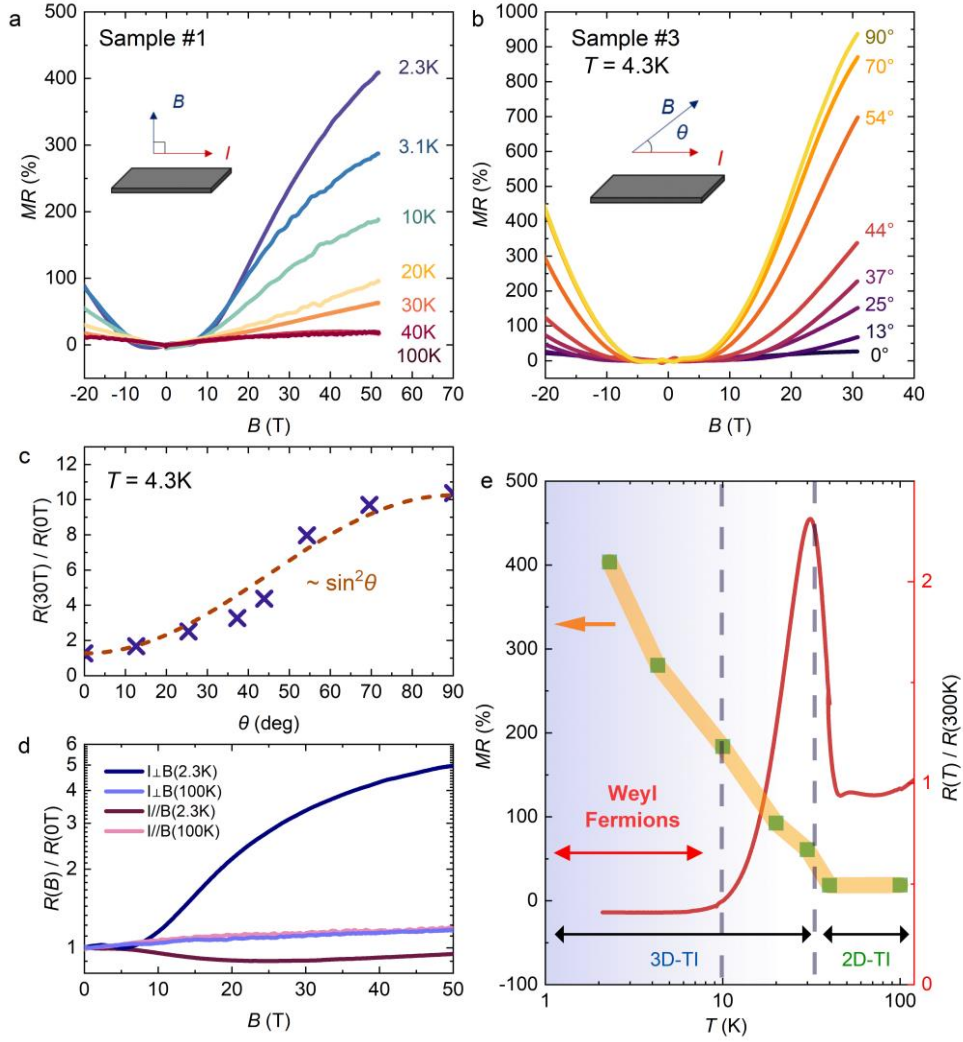

**Figure 3 | Positive and non-saturating MR in α-(BETS)$_2$I$_3$. a**, Transverse MR of α-(BETS)$_2$I$_3$ ($I \perp B$). The magnetic-field dependence of MR of Sample #1 measured at several temperatures. a large positive and non-saturating MR can be observed. **b**, The magnetic field dependence of MR of Sample #3 at 4.3 K measured at several angles $\theta$. The angle $\theta$ is defined by the angle between $I$ and $B$. **c**, Angle dependence of MR of Sample #3 at 30T. The orange dashed line is the fitting line of $\sin^2\theta$. **d**, Comparison of MRs of Sample #1 of $I \perp B$ and $I // B$ at 2.3 and 100 K, respectively. The strong anisotropy of MR can be observed only at low temperatures. **e**, Temperature dependence of the

magnitude of the transvers MR of Sample #1 and the out-of-plane resistance of Sample #2. Below 35 K, the positive MR begins to increase, which agrees with the appearance of the surface conduction on the *ab*-plane.

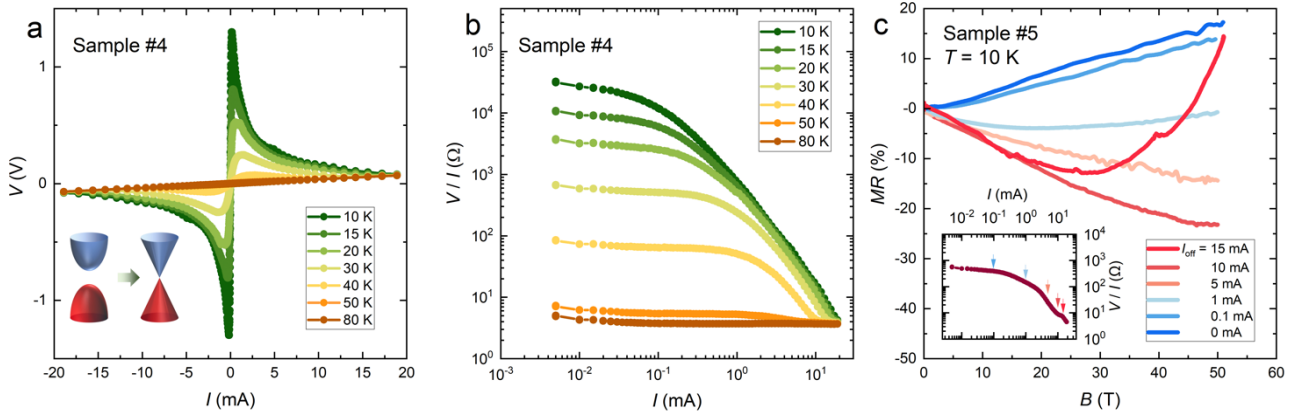

**Figure 4 | Current switching of the topological phase in α-(BETS)$_2$I$_3$. a,** Current−voltage ($I-V$) characteristic of Sample #4 measured at several temperatures using the four-terminal dc method with constant currents. A nonlinear conduction effect is observed in the wide current range. **b,** Current dependence of the sample resistance ($V/I$). In the low current region, $V/I$ is a constant value followed by the Ohm's law. As the applied current increases, the sample resistance decreases by several orders of magnitude. **c,** Longitudinal MR of Sample #5 ($I // B$) at 10 K measured under the several values of the offset dc current ($I_{off}$). The positive to negative change of MR indicates the topological phase switching from TI to the Dirac semimetal state. Inset: the $I$ - $V/I$ character of sample #5. Nonlinear conduction is clearly observed above $I > 0.1$ mA.

**Methods**

Single crystals of α-(BETS)$_2$I$_3$ were grown using a standard electrochemical method. The in-plane and out-of-plane resistances as shown in Fig. 2 were measured using the standard four-terminal method in a physical property measurement system (Quantum Design). To perform the in-plane resistance and magnetoresistance (MR) measurements, four gold wires were attached to the conducting plane (*ab*-plane) using a carbon paste. Dimensions of measured Samples #1–#5 are summarized in Tab. S1. Because the crystals were very thin, there is considerable ambiguity in thickness *t*. In the case of out-of-plane measurements, two gold wires were attached to the top of the *ab*-plane, whereas the other two were attached to the back.

The MR measurements in pulsed magnetic fields were performed using the standard ac four-terminal method with an excitation voltage at a frequency of 2–5 kHz. The amplitude of excitation voltage was set within the ohmic conduction range of the samples. Regarding the low-noise ac resistance measurements, the voltage signal was amplified using pre-amplifiers (Model SR560, Stanford Research Systems). Furthermore, the angle of *B* was calculated as the ratio of induced voltages of multiple pickup coils, and temperature was measured using a calibrated Cernox thermometer and a temperature controller (Lakeshore model 335).

The current–voltage (*I–V*) characteristics were measured using a short pulse current of 10 ms produced by a current source (NI-9265, National Instruments) and a multimeter (Model 2000, Keithley). To avoid the self-heating effect, this measurement was performed under exchange gas conditions.

The current dependence of the longitudinal MR was measured using the four-terminal method with an ac excitation and a dc offset current. The dc offset current ($I_{\text{off}}$) was produced by the current source NI-9265. The amplitude of the ac current was set to approximately 10% of the magnitude of $I_{\text{off}}$ to satisfactorily avoid the change of electronic state by an ac component. Although the application of the excitation ac voltage induced a mismatch between the actual sample resistance and estimated value from the $I$–$(V/I)$ characteristic, a combination of dc and ac did not have extrinsic effects on the MR properties. Therefore, the effect of the current-induced electronic state on the magnetic response could be investigated.

**Supplemental Materials for**
**"Correlation-driven Organic 3D Topological Insulator with Relativistic Fermions"**

Tetsuya Nomoto[1,2,*], Shusaku Imajo[1], Hiroki Akutsu[2], Yasuhiro Nakazawa[2], Yoshimitsu Kohama[1]

[1]*The Institute for Solid State Physics, the University of Tokyo, Kashiwa, Chiba 277-8581, Japan*
[2]*Graduate School of Science, Osaka University, Toyonaka, Osaka 560-0043, Japan*

**Table of Contents**

## §1 Anisotropy of resistivity of α-(BETS)$_2$I$_3$

In this section, the reason for the out-of-plane resistance clearly detecting the formation of the metallic surface compared to the in-plane resistance is addressed. Because a three-dimensional (3D) topological insulator (TI) has bulk and surface conduction channels, the observed resistance is a combined resistance of the bulk resistance ($R_{bulk}$) and the surface resistance ($R_{surface}$). If $R_{bulk}$ is much larger than $R_{surface}$, then the contribution of $R_{surface}$ to the combined resistance is dominant, and consequently, the resistance significantly changes when the surface conduction appears. In α-(BETS)$_2$I$_3$, the in- and out-of-plane resistivities, which correspond to $R_{bulk}$ at room temperature, are $4.6\times10^{-2}$ Ω cm and $1.2\times10^{1}$ Ω cm at 300 K, respectively; the out-of-plane resistivity is approximately 1000 times larger than the in-plane resistivity. Because of the high out-of-plane resistivity of the bulk region, it is reasonable that the influence of the surface metallic state is clearly observed in the out-of-plane configuration as shown in Fig. 2. Although a clear drop in resistance was not observed in the in-plane configuration, the deviation from the Arrhenius law and saturating behaviour of resistance below 35 K were detected, suggesting the emergence of a metallic surface state. Moreover, no in-plane anisotropy was observed in the experiments.

## §2 Sample dependence and reproductivity of resistance

In this section, the reproducibility of the temperature dependence of the resistance is discussed. In a normal bulk insulator or bulk metal, the resistance is inversely proportional to the sample size. Thus, the resistivity or normalised resistance by the value at room temperature does not depend on the sample size. In contrast, when the resistivity of the bulk and surface is different as in a TI, the resistance is determined by the ratio of the bulk and surface conductance. Because the thickness of the surface conducting layer depends on the sample, the volume fraction changes with the sample size. Therefore, the resistance of TIs cannot be simply normalised by the sample size or room-temperature resistance. The sample size dependence of the resistivity is often observed in 3D-TI [1-4].

Figure S1 shows the Arrhenius plot of the in-plane resistance of nine different samples, including Sample #1 (Fig. 2) and Sample I1–I8. Above 35 K, all the data can be scaled on a single curve, suggesting the absence of surface conduction. In contrast, below 35 K, the sample dependence can be clearly observed, which strongly suggests the presence of surface conduction below 35 K.

In the case of the out-of-plane resistance, the sample dependence is more complicated because of a stray current, that is, a non-negligible contribution from the in-plane resistance exists. Although the resistance drop at approximately 35 K can be reproduced in a different sample (Sample O1), as shown in Fig. S2A, several samples did not show a clear resistance drop due to the stray current, as shown in Fig. S2B (Sample O2). However, a plateau region of the resistance between 10 and 35 K and an increase of resistance below 10 K were observed for all samples. These behaviours can be reasonably explained if the bulk conduction is dominant, where the contribution from a small portion of surface conduction is negligible. Moreover, negative MR was not observed in the out-of-plane resistance measurement, even in the case of $I \parallel B$, as shown in Fig. S3. This is reasonable because the direction of the voltage ($I \parallel c$-axis) is different from that of the surface conduction flow.

### §3 Magnetoresistance in the two-carrier model

In this section, the two-carrier model mentioned in the main text is explained. In the classical model [5], the MR $\Delta R / R$ (0 T) of the two carrier systems is described by the following equation:

$$\frac{\Delta R}{R(0T)} = \frac{r_e r_h (R_e + R_h) + (r_e R_h^2 + r_h R_e^2) B^2}{(r_e + r_h)^2 + (R_e + R_h)^2 B^2} \times \frac{r_e + r_h}{r_e r_h} - 1 \qquad (1),$$

where $r_e$ and $r_h$ are the resistances of electrons and holes, respectively, and $R_e$ and $R_h$ are the Hall coefficients of the electron and hole carriers, respectively. Because the two carriers mutually depress the Hall electric field formed by the polarisation owing to the Lorentz force, MR becomes large and does not saturate until high magnetic fields. This model can explain the large MR value of materials with a semimetallic band such as graphene, Bi, and some topological semimetals [6-9].

Figure S4 shows MR when $B$ is perpendicular to $I$ (blue series) and the fitting curve using Eq. 1. In this analysis, it was assumed that $r_n(T) = r_h(T) (= r_0(T))$. The equation can reproduce MR quantitatively, indicating that the anomalously large MR of $\alpha$-(BETS)$_2$I$_3$ is induced by the Dirac-cone-type semimetallic band structure. However, the minor mismatch of the fitting below 20 T is probably because of the negative MR caused by the longitudinal component.

## §4 Non-equilibrium metallic state in $\alpha$-(BETS)$_2$I$_3$

In this section, we supplement the non-linear conduction and current-induced metallic state in $\alpha$-(BETS)$_2$I$_3$. As mentioned in the main text, the current–voltage ($I$–$V$) curves exhibit peak structures at $V_{peak}$ (Fig. S5a). The relationship between $V_{peak}$ and $T$ is shown in Fig. S5b. The $V_{peak}$ exponentially increases with decreasing temperature, and thus it can be fitted well using $V_{peak} = V_0 \exp(-T/T_0)$, where $V_0$ and $T_0$ are the fitting parameters. Using the characteristic values $V_0 = 3.18$ V ($= 3.18$ kV/cm) and $T_0 = 11.1$ K, we succeeded in fitting our results. This relation is often observed in materials that exhibit nonlinear conduction induced by the electron delocalization, such as charge-density-wave and Mott insulators. The exponential increase of $V_{\text{peak}}$ indicates that the nonlinear conduction of α-(BETS)$_2$I$_3$ cannot be explained by the classical models where the threshold voltage weakly depends on temperature.

In the main text, we speculated that the suppression of the bandgap on the Dirac band is the origin of the giant nonlinear conduction. Thus, we measured the excitation current dependence of the in-plane resistance using the four-terminal method to verify this speculation. Fig. S6 shows the temperature dependence of the in-plane resistance of Sample S1 measured by applying several external current values. For small excitation currents, the temperature dependence of the resistance reproduces the result of Sample #1 as shown in Fig. 1d. As the excitation current is increased, the increase of resistance below the metal-insulator transition becomes slower. Finally, a temperature-independent resistance is observed over a wide temperature range above approximately 10 mA. Such temperature-independent resistance is a characteristic feature of zero-gap semimetals in organic conductors [10, 11]. Using the Arrhenius plot from the data between 30–40 K, as illustrated in the inset of Fig. S6, we calculated the gap size $\Delta$ when the excitation currents were applied. Indeed, $\Delta$ decreases with increasing current value and asymptotically approaches zero. Consequently, the current-induced metallic state is a zero-gap Dirac semimetal.

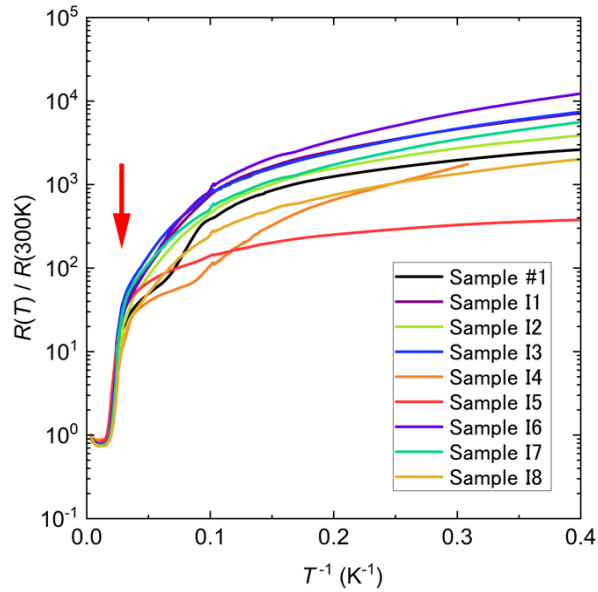

**Figure S1 | Arrhenius plot of the in-plane resistance on α-(BETS)$_2$I$_3$ for Sample #1 and Sample I1-I8.** The resistance is normalised by the value at room temperature. The red arrow corresponds to the crossover temperature $T$ = 35 K. Below 35 K, the sample dependence of resistance can be clearly observed, indicating the surface conduction in low temperatures.

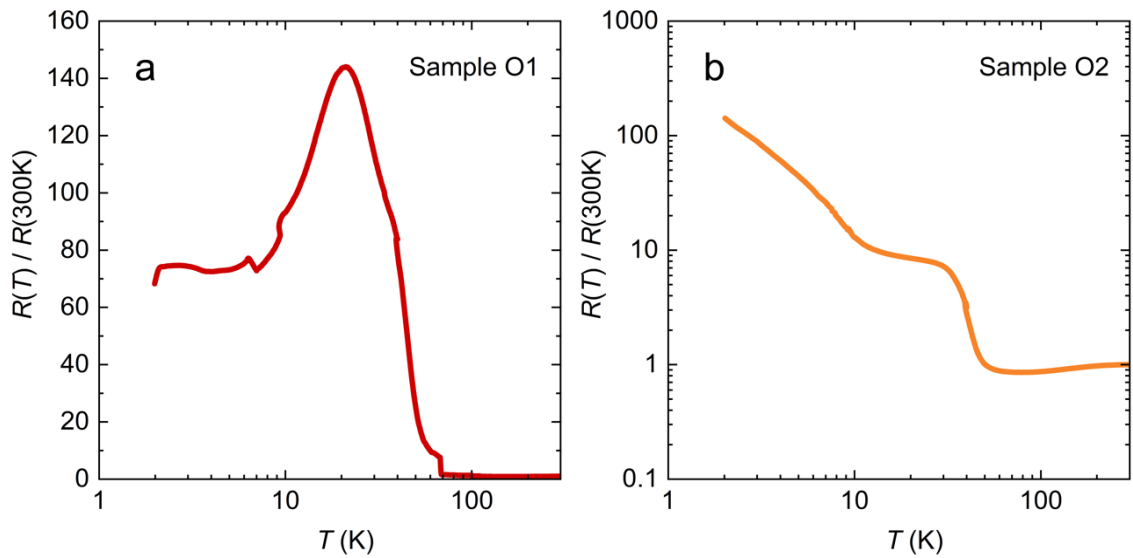

**Figure S2 | Sample dependence of the out of plane resistance on α-(BETS)$_2$I$_3$. a,** Example of the sample showing the resistance drop at approximately 30 K (Sample O1). **b,** Example of the sample not showing the resistance drop at approximately 35 K (Sample O2). A plateau of the resistance was observed between 35 and 10 K, indicating the existence of the surface metallic state.

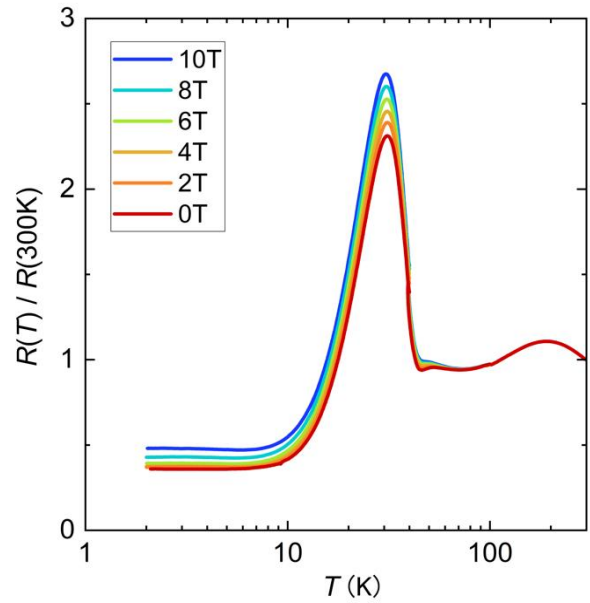

**Figure S3 | Temperature dependence of the out-of-plane resistance (Sample #2) under magnetic fields of 0, 2, 4, 6, 8, and 10 T.** The magnetic field was applied perpendicular to the *ab*-plane ($B \parallel c$-axis). The resistance monotonically increases with increasing magnetic fields. A negative MR cannot be observed in the out-of-plane resistance measurements.

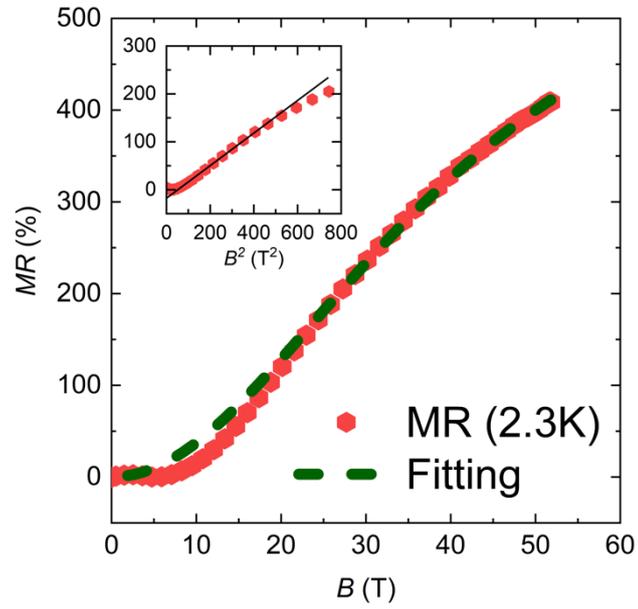

**Figure S4 | Positive MR at 2.3 K under the $I \perp B$ configuration in Sample #1.** The green dashed line is the fitting curve using equation (1). The parameters used in the fit are $r_0 = 1.31 \times 10^{-3}$, $R_h = 5.05 \times 10^{-5}$, and $R_e = 1.14 \times 10^{-4}$. Inset: MR vs $B^2$ plot. In the weak magnetic field region below 20 T, MR follows $B^2$ (the black solid line).

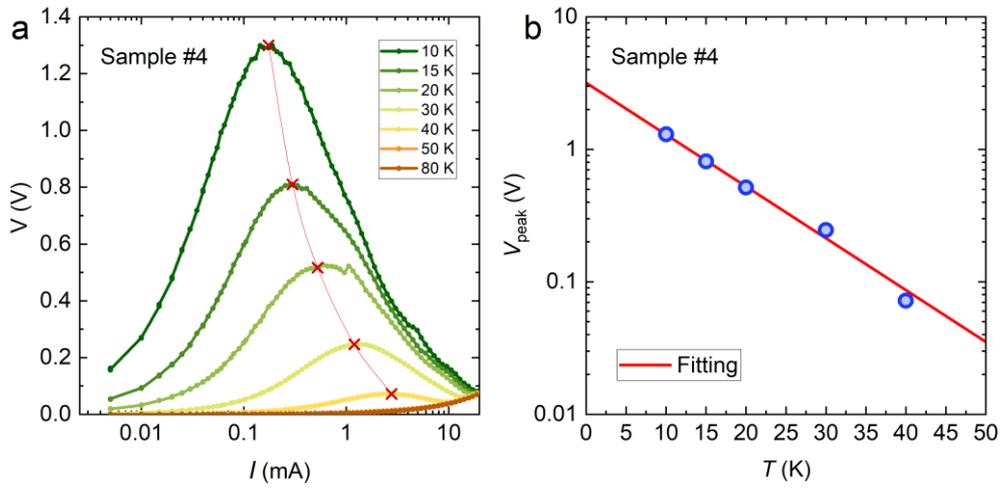

**Figure S5 | Current-voltage character of α-(BETS)$_2$I$_3$. a**, Enlarged figure of the current-voltage characteristic of Sample #4. The red dashed line points out $V_{peak}$. **b**, $V_{peak}$ versus $T$. The red solid line is the fitting curve $V_{peak}(T) = V_0\exp(-T/T_0)$ using $V_0 = 3.18$ V and $T_0 = 11.1$ K.

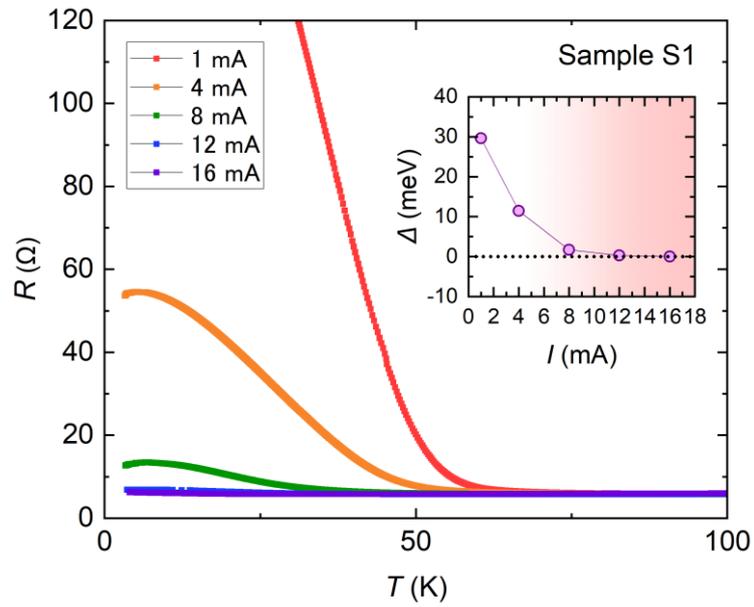

**Figure S6 | Temperature dependence of the in-plane resistance of α-(BETS)$_2$I$_3$ (Sample S1) measured by using a variety of excitation currents.** The temperature-independent resistance is observed in the excitation currents of 12 and 16 mA, which is a typical feature of zero-gap systems. Inset: current dependence of the gap size estimated by each resistance curve.

**Table S1 | Dimensions (length, width, and thickness) of the measured samples #1~#5.**

| Sample name | Length (μm) | Width (μm) | Thickness (μm) |
|---|---|---|---|
| #1 | $2.0\times10^2$ | $1.8\times10^3$ | $5.0\times10^1$ |
| #2 | $1.0\times10^3$ | $5.6\times10^2$ | $1.0\times10^1$ |
| #3 | $1.6\times10^2$ | $7.2\times10^2$ | $3.0\times10^1$ |
| #4 | $1.4\times10^2$ | $7.2\times10^2$ | $1.0\times10^1$ |
| #5 | $1.0\times10^2$ | $9.0\times10^2$ | $4.0\times10^0$ |